\documentclass[letter]{aa}  

\usepackage{graphicx}
\usepackage{txfonts}
\usepackage{epsf,epsfig}

  \def\nH{n_{\langle{\rm H}\rangle}}
  \def\Tg{T_{\rm g}}
  \def\Td{T_{\rm d}}
  \def\etal{${\rm \hspace*{0.8ex}et\hspace*{0.7ex}al.\hspace*{0.65ex}}$}
  \def\plus{${\rm \hspace*{0.7ex}\&\hspace*{0.7ex}}$}
  \def\ie{i.\,e.\ }
  \def\eg{e.\,g.\ }

\begin{document}

\title{Too little radiation pressure on dust\\[0.05ex] 
       in the winds of oxygen-rich AGB stars}


\titlerunning{Too little radiation pressure on dust
       in winds of oxygen-rich AGB stars}

\author{Peter Woitke}

\institute{Sterrewacht Leiden, P.O.~Box 9513, 2300 RA Leiden, The Netherlands,
           \email woitke@strw.leidenuniv.nl }

\date{Received 30 August 2006; accepted date}

\abstract
  {}
  {{It is commonly assumed that the massive winds of AGB stars are
   dust-driven and pulsation-enhanced. However, detailed
   frequency-dependent dynamical models which can explain the observed
   magnitudes of mass loss rates and outflow velocities have
   been published so far only for C-stars. This letter reports
   on first results of similar models for oxygen-rich AGB stars. The
   aim is to provide a better understanding of the wind driving
   mechanism, the dust condensation sequence, and the role of
   pulsations.}}
  {{New dynamical models for dust-driven winds of oxygen-rich AGB stars
   are presented which include frequency-dependent Monte Carlo
   radiative transfer by means of a sparse opacity distribution
   technique and a time-dependent treatment of the nucleation, growth
   and evaporation of inhomogeneous dust grains composed of a mixture of}
   Mg$_2$SiO$_4$, SiO$_2$, Al$_2$O$_3$, TiO$_2$, and solid Fe.}
  {{The frequency-dependent treatment of radiative transfer reveals
   that the gas is cold close to the star ($700-900\,$K at
   $1.5-2\,R_\star$) which facilitates the nucleation process. The
   dust temperatures are strongly material-dependent, with differences 
   as large as 1000\,K for different
   pure materials, which has an important influence on the dust formation
   sequence. Two dust layers are formed in the dynamical models:
   almost pure glassy Al$_2$O$_3$ close to the star
   ($r\!\ga\!1.5\,R_\star$) and the more opaque Fe-poor Mg-Fe-silicates further
   out. Solid Fe or Fe-rich silicates are found to be the only 
   condensates that can efficiently absorb the stellar light in the near IR. 
   Consequently, they play a crucial role for the wind driving mechanism and 
   act as thermostat. Only small amounts of Fe can be
   incorporated into the grains, because otherwise the grains get too hot.
   Thus, the models reveal almost no mass loss, and no dust shells.}}
  {The observed dust sequence Al$_2$O$_3$ $\to$ Fe-poor Mg-Fe-silicates for
   oxygen-rich AGB stars having low $\to$ high mass loss rates is in
   agreement with the presented model and can be understood as
   follows: Al$_2$O$_3$ is present in the extended atmosphere of the
   star below the wind acceleration region, also without mass loss.
   The Mg-Fe-silicates form further out and, therefore, their amount 
   depends on the mass loss rate. The driving mechanism of oxygen-rich 
   AGB stars is still an unsolved puzzle.}


\maketitle
%

\section{Introduction}

The mass loss mechanism of AGB stars and red supergiants is a
long-standing astrophysical problem.  In the carbon-rich case, an
extraordinary condensate exists (amorphous carbon) which is very
stable, \ie it can exist already close the star, and is very opaque in
the optical and near IR spectral region.  Detailed dynamical models
with time-dependent dust formation theory (Winters\etal
2000)\nocite{wlj2000} show that the formation of amorphous carbon can
provide sufficient radiation pressure to drive massive outflows,
consistent with the basic characteristics of C-star winds. This result has
been confirmed by dynamical models with frequency-dependent radiative
transfer by H{\"o}fner\etal(2003)\nocite{hga2003}.

However, in the oxygen-rich case, no such condensate exists. The most
stable metal-oxides like Al$_2$O$_3$ are too rare. The abundant pure
silicates like Mg$_2$SiO$_4$ are already less stable and almost
completely transparent around $1\,\mu$m where most of the stellar flux
escapes.  Solid Fe and Mg-Fe-silicates are opaque but even less
stable. Stationary models of dust-driven O-rich AGB star winds with
grey radiative transfer (Ferrarotti\plus Gail 2006) and dynamical
models with pulsation and grey radiative transfer (Jeong\etal 2003)
nevertheless came to the conclusion that the winds of O-rich AGB stars
are dust-driven, where the stellar pulsation helps to provide the
necessary density conditions to form the dust close to the star
(``pulsation-enhanced'').

In contrast, the a posteori frequency-dependent radiative transfer
analysis of non-linear pulsation models with simplified dust
formation theory by Ireland\plus Scholz (2006) did not find much
radiation pressure on dust (Al$_2$O$_3$ and
Mg$_{2x}$Fe$_{2-2x}$SiO$_4$) in O-rich Mira variables, with radiative
accelerations as small as 0.08 to 0.29 times the gravitational
deceleration.

Recent mid-IR observations of O-rich AGB stars in globular
clusters with {\sc Spitzer} (Lebzelter\etal 2006) and of galactic
bulge AGB stars with ISO (Blommaert et al.\ 2006) show a clear
correlation between the kind of condensate and the mass loss rate
$\dot{M}$, called the ``observational dust condensation sequence'': stars
with low $\dot{M}$ show mainly Al$_2$O$_3$, whereas stars with higher
$\dot{M}$ show increasing amounts of Mg-Fe-silicates.  From {\sc Midi}
interferometry of the red supergiant $\alpha$\,Ori,
Verhoelst et al.\ (2006)\nocite{vdmhc2006} concluded that Al$_2$O$_3$ grains 
are already present at radial distances as small as 
$r\!=\!1.5\,R_\star$.

\section{The model}

\noindent{\bf Hydrodynamics} is solved by using the {\sc
Flash}-solver (Fryxell et al.\ 2000)\nocite{for2000} in spherical
symmetry, including gravity and self-developed modules for radiation
pressure on dust\plus molecules and radiative heating\,/\,cooling
(see Woitke 2006a for details)\nocite{woi2006a}. We use
here, however, a piston approximation as inner boundary condition to
simulate the pulsation of the star, and a new equation of state for a
mixture of H$^+$, e$^-$, H, H$_2$, He and other atomic metals in LTE,
including ionisation and dissociation potentials, and vibrational
and rotational excitation energies of H$_2$.

\smallskip
\noindent{\bf Radiative transfer:} For the models presented in this
letter, we have developed a new frequency-dependent Monte Carlo
radiative transfer technique which allows for arbitrarily high optical
depths (Woitke 2006b)\nocite{woi2006b}. In frequency-dependent
stellar atmospheres, the gas is always optically thick at least in some
wavelengths ($\sim\!10^{\,5}$), which is a big problem for standard MC
techniques. The method is coupled to the hydrodynamics and can be used
also for 2D models.

\smallskip
\noindent{\bf Opacities:} The basis for our radiative transfer
treatment are monochromatic molecular gas opacities from the {\sc
Marcs} stellar atmosphere code (J{\o}rgensen\etal 1992)\nocite{j92},
extracted by Helling\etal (2000)\nocite{hws2000}. The frequency space
is subdivided into five spectral bands with two opacity distribution
points in each band, resulting in altogether $5 \times 2$ effective
wavelengths sampling points (for the traditional ODF
approximation see \eg Carbon 1979)\nocite{car79}. High and low mean opacity
values are tabulated for each spectral band during the initialisation
phase of the program in such a way that they simultaneously result in
the correct Planck {\it and} Rosseland band-mean gas opacities. The
details will be explained in another paper (Woitke 2006b).  

\begin{figure}
  \epsfig{file=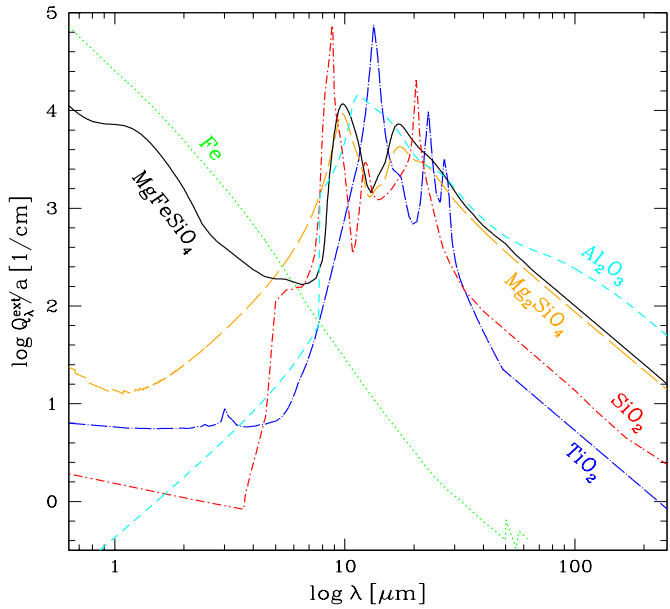,width=8.8cm,height=8cm}
  \caption{Extinction efficiencies over particle radius $Q_{\rm ext}/a$
           in the Rayleigh-limit of Mie theory according to the Jena optical 
           data base. The data is partly log-log extrapolated.}
  \vspace*{-1mm}
  \label{fig:kappa_dust}
\end{figure}

Dust opacities are calculated in the Rayleigh limit of Mie theory
according to the Jena optical data base, kindly provided by Th.~Posch
(see Fig.~\ref{fig:kappa_dust}). The total dust extinction coefficient
$\rm[cm^2/g]$ is assumed to be given by\footnote{According to our
assumption of dirty grains, an application of effective medium theory
would be more appropriate, which will be examined in a future
paper. First results show that the effective extinction is stronger
than the simple volume-means used in this paper.}
\begin{equation}
  \hat{\kappa}_{\lambda,\rm\,ext}^{\rm\;dust} 
  = \frac{3}{4}\,L_3 \sum_s \frac{V_s}{V_{\rm tot}} Q^s_{\rm ext}(a,\lambda)/a
\end{equation}
where $L_3$ is the third moment of the dust size distribution function
and $V_s/V_{\rm tot}$ is the volume fraction of solid material $s$ in the
dust component. The extinction efficiencies over particle radius of the pure
solids $s$ are shown in Fig.~\ref{fig:kappa_dust}. Note that most
oxygen-rich condensates have a ``glassy'' character. They are almost
transparent in the optical and near IR but opaque in the mid IR where
the strong vibrational resonances are situated. In contrast, solid Fe
and Fe-rich silicates are opaque even in the optical. The
monochromatic dust opacities are subject to the same averaging
procedure to result in high/low band-mean dust opacities as described
above for the gas opacities.

\smallskip
\noindent{\bf Dust formation} is described by a system of differential
moment equation explained in (Helling\plus Woitke 2006)\nocite{hw2006}
considering the growth and evaporation of inhomogeneous dust grains
composed of a mixture of Mg$_2$SiO$_4$, SiO$_2$, Al$_2$O$_3$, TiO$_2$,
and solid Fe (13 growth/evaporation reactions). The nucleation rate of
$\rm(TiO_2)_N$ clusters is adopted from Jeong
(2000)\nocite{jeo2000}. The molecular concentrations entering into the
calculation of the nucleation and growth rates are calculated by a
small neutral equilibrium chemistry for 11 atoms (H, He, C, O, N, Mg,
Al, Si, S, Ti, Fe) and 33 molecules (H2, CO, CO$_2$, OH, H$_2$O,
CH$_4$, N$_2$, CN, HCN, NH$_3$, H$_2$S, SiS, SO, HS, SiO, SiH,
SiH$_4$, SiO$_2$, SiN, SO$_2$, MgH, MgS, MgO, MgOH, Mg(OH)$_2$, FeO,
Fe(OH)$_2$, AlOH, AlO$_2$H, Al$_2$O, AlH, TiO, TiO$_2$).

\section{The static solution}
\label{sec:static}

\begin{figure}
  \hspace*{-1mm}\epsfig{file=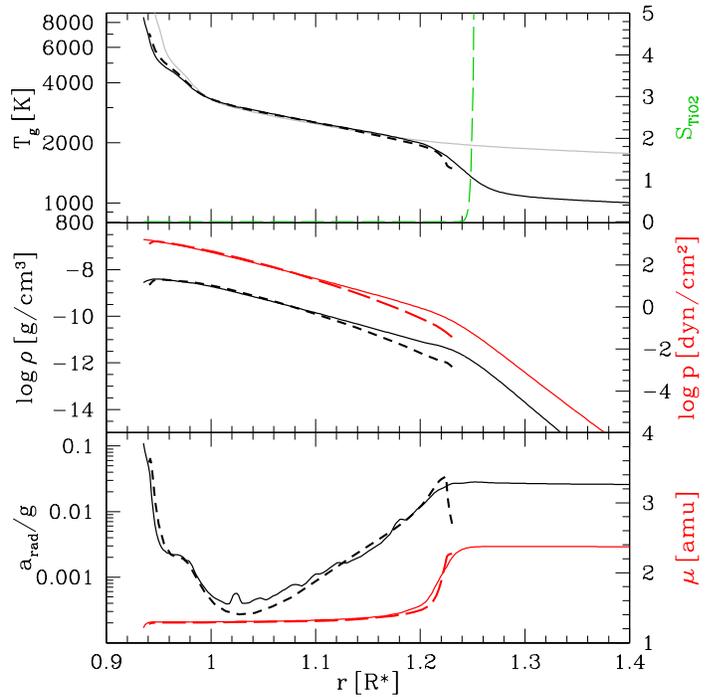,width=9.2cm}
  \caption{Hydrostatic initial model (full) in
           comparison to a spherical {\sc Marcs} model
           (dashed) and a grey model (grey $\Tg$-line) for 
           $M_{\star}\!=\!1\,M_{\odot}$,
           $T_{\star}\!=\!2800\,$K, $\log g\!=\!-0.6$
           ($L_{\star}\!=\!6048\,L_{\odot}$), $Z\!=\!1$. $S_{\rm{\!TiO2}}$
           shows the supersaturation ratio of $\rm TiO_2$, indicating that
           nucleation is already possible very close to the star. 
           Long-dashed graphs belong to the r.h.s.\ axis.}
  \vspace*{-1mm}
  \label{fig:Tstruc}
\end{figure}

The hydrostatic, dust-free solution (see Woitke 2006a for equations
and further details) used as initial model for the dynamical models is
shown in Fig.~\ref{fig:Tstruc}. The obtained degree of agreement with
the {\sc Marcs} model is quite remarkable for such a rough
10-wavelengths-point treatment of radiative transfer. Similar results
have been obtained by H{\"o}fner\etal(2003) and Helling\plus
J{\o}rgensen (1998)\nocite{hj98}, using a sparse opacity sampling
technique with about 100 sampling points. The gas temperature
structure $\Tg(r)$ shows a pronounced ``step'' of almost 1000\,K
around $1.25\,R_\star$ (the {\sc Marcs} model is actually not extended
enough to reveal this step completely). It is clear, however, that
this is not just a ``surface effect''. Outside of this $\Tg$-step even
the strongest molecular lines become optically thin (in the
hydrostatic case) and the line blanketing effect works at full
strength. The grey model fails completely in predicting this step
which is very meaningful for the dust formation. The agreement with
the {\sc Marcs} results concerning the mass density $\rho(r)$, the gas
pressure $p(r)$, the mean molecular weight $\mu(r)$, and the
acceleration by radiation pressure on molecules divided by gravity
$\Gamma_{\rm gas}=a_{\rm rad}/g(r)$ is also satisfying.

\section{Rough estimates of the dust acceleration}

The calculation of the spectral mean intensities $J_{\lambda\,}(r)$ and
spectral fluxes $F_{\!\lambda\,}(r)$ in the static model (see
Fig.~\ref{fig:Tstruc}) allows for a quick estimate of the maximum
possible radiative acceleration by dust if the dust is still optically
thin. Table~\ref{tab:dust_estimates} shows the resulting dust
temperatures $\Td$ of several pure condensates
\begin{equation}
    \int\!\hat{\kappa}^{\rm\;dust}_{\lambda,\rm\,abs} 
                  \,J_{\lambda\,}(r) \,d\lambda 
  \;= \int\!\hat{\kappa}^{\rm\;dust}_{\lambda,\rm\,abs} 
                  \,B_\lambda(\Td) \,d\lambda 
\end{equation}
and the radiative acceleration by dust divided by gravity
\begin{equation}
  \Gamma_{\rm dust}(r) \,=\, 
  \frac{\frac{1}{c} \int \hat{\kappa}^{\rm\;dust}_{\lambda,\rm\,ext}
                          \,F_{\!\lambda\,}(r) \,d\lambda}
       {\frac{G M(r)}{r^2}}
\end{equation}
at three selected distances from the star. For each condensate
we take the maximum possible dust volume per mass 
$L_{3,\rm\,max}$\linebreak\\*[-2.3ex]
given by element conservation constraints, \eg
$L_{3,\rm\,max}^{\rm Mg_2SiO_4}=$\linebreak\\*[-2.3ex]
${\rm Min}\big\{\frac{1}{2}\epsilon_{\rm Mg},\epsilon_{\rm Si},
      \frac{1}{4}\epsilon_{\rm O}\big\}\,V_{\rm Mg_2SiO_4}\,\nH/\rho$
where $\nH$ is the hydrogen nuclei density, $\epsilon_k$ the abundance 
of element $k$ and $V_{\rm Mg_2SiO_4}$ the monomer volume of Mg$_2$SiO$_4$
(see Helling\plus Woitke 2006)\nocite{hw2006}.

The results shown in Table~\ref{tab:dust_estimates} demonstrate that
the dust temperatures $\Td$ are strongly material-dependent, with
differences as large as 1000\,K at the same distance from the star,
which is a remarkable result. All condensates (except solid Fe) have
strongly peaked mid-IR resonances which are situated just around the
maximum of the local Planck function -- they work perfect for
radiative cooling. In contrast, the glassy character of the oxides and
pure silicates like Al$_2$O$_3$, SiO$_2$, Mg$_2$SiO$_4$ and MgSiO$_3$
(the low absorption efficiencies at optical and near-IR wavelengths,
see Fig.~\ref{fig:kappa_dust}) prevent efficient heating by the
star. Consequently, the pure glassy condensates can exist
astonishingly close to the star (see also Woitke 1999).\nocite{woi99}

\begin{table}
\caption{Calculated dust temperatures $\Td$ (first row) and dust radiative 
         accelerations $\Gamma_{\rm dust}=a^{\rm dust}_{\rm rad}/g$ 
         (second row) in case of full condensation of several
         condensates into small particles
         in the static model (see Fig.~\ref{fig:Tstruc}). The resulting
         dust-to-gas ratio $\rho_{\rm dust}/\rho_{\rm gas}$ is shown
         in the middle column. Temperatures values with $^\star$ mark 
         thermally unstable condensates.}
\label{tab:dust_estimates}
\begin{tabular}{c|c|ccc}
  solid material
  & $\frac{\rho_{\rm dust}}{\rho_{\rm gas}}\,[10^{-3}]$
  & $r=1.5\,R_\star$ 
  & $r=2\,R_\star$ 
  & $r=5\,R_\star$ \\
\hline
 & & & & \\*[-1.7ex]
{\sf TiO$_2$}          & 0.0061
                            & 1030\,K 
                            &  750\,K 
                            &  380\,K \\ 
                       &  & 0.00004
                          & 0.00004
                          & 0.00005 \\
{\sf Al$_2$O$_3$}      & 0.11
                            & 1090\,K 
                            &  810\,K 
                            &  420\,K \\
                       &  & 0.0013
                          & 0.0014
                          & 0.0015  \\
{\sf SiO$_2$}          & 1.6
                            & 1000\,K 
                            &  740\,K 
                            &  380\,K \\
                       &  & 0.032
                          & 0.034
                          & 0.036   \\
{\sf Mg$_2$SiO$_4$}    & 1.9
                            & 1150\,K 
                            &  850\,K 
                            &  430\,K \\
                       &  & 0.022
                          & 0.024
                          & 0.025   \\
{\sf MgFeSiO$_4$}      & 4.0
                            & 1930\,K$^\star$ 
                            & 1710\,K$^\star$ 
                            & {\bf 1170\,K} \\
                       &  & {1.3}
                          & {1.4}
                          & {\bf 1.4}      \\
{\sf MgSiO$_3$}        & {2.3} 
                            & {1010\,K} 
                            & { 740\,K} 
                            & { 380\,K} \\
                       &  & {0.025}
                          & {0.027}
                          & {0.029}   \\
{\sf Mg$_{\,0.5}$Fe$_{\,0.5}$SiO$_3$} & {3.0} 
                            & {1880\,K}$^\star$  
                            & {1580\,K}$^\star$  
                            & {690\,K}  \\
                       &  & {0.21}
                          & {0.21}
                          & {0.18}    \\
{\sf Fe}               & {1.3} 
                            & {1980\,K}$^\star$   
                            & {1770\,K}$^\star$  
                            & {\bf 1280\,K} \\
                       &  & {0.85}
                          & {0.89}
                          & {\bf 0.88}     \\
{\sf am.\,carbon}      & {3.0}
                            & {1870\,K}$^\star$   
                            & {\bf 1640\,K} 
                            & {\bf 1130\,K} \\
 (C/O\,=\,1.5)         &  & {20}
                          & {\bf 21}
                          & {\bf 21}       
\end{tabular}
\end{table}

For the same reasons, radiative pressure on all glassy condensates is
negligible!  It is without effect for the wind acceleration
mechanism whether for example Mg$_2$SiO$_4$ condenses out or not.  The
only dust species that can potentially drive a stellar outflow are
solid Fe and Fe-rich silicates like MgFeSiO$_4$. 



The unavoidable consequence of the spectral characteristics of oxygen-rich
dust is that radiative acceleration must be paid for by radiative
heating, \ie dust species capable of driving a stellar wind
($\Gamma\!\ga\!1$) are hot and can only exist at a relatively large
distance from the star (\eg $r\!\ga\!5\,R_\star$, marked in bold in
Table~\ref{tab:dust_estimates}). For comparison, amorphous carbon is so
opaque and stable (in a C-rich gas) that it could accelerate the gas
already from $2\,R_\star$ onwards, with $20\,\times$ the local gravity
in this model.

\section{Results of the dynamical models}

\begin{figure}
  \epsfig{file=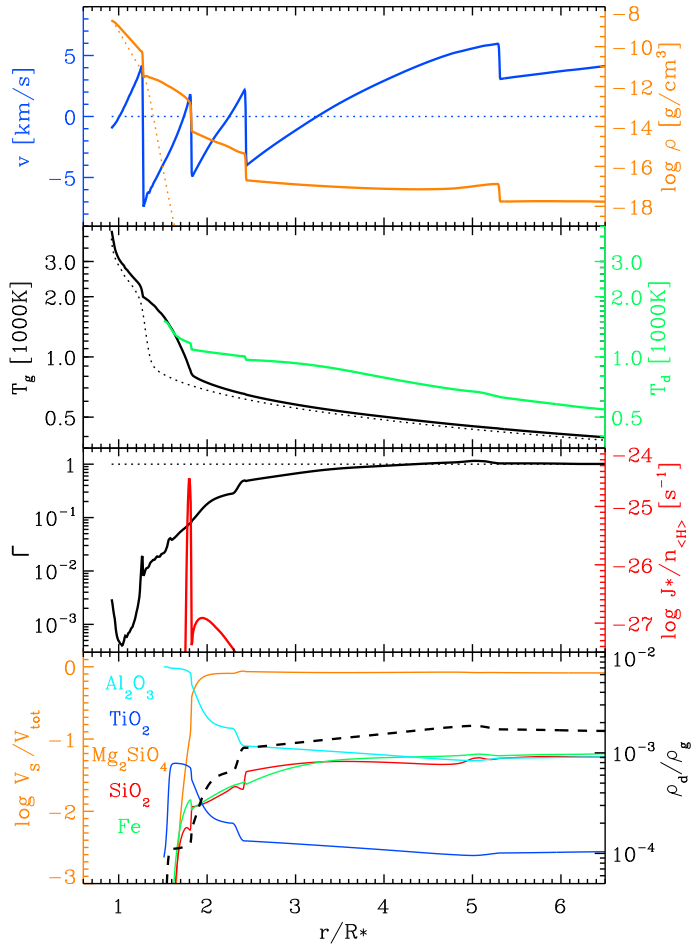,width=9.1cm}
  \caption{Dynamical wind model after 100 years of simulation time.
           Parameter: $M_{\star}\!=\!1\,M_{\odot}$,
           $T_{\star}\!=\!2500\,$K, $L_{\star}\!=\!10000\,L_{\odot}$
           ($\log g\!=\!-1.015$), $Z\!=\!1$, piston period $P\!=\!600$
           days and velocity amplitude $\Delta{\rm
           v}\!=\!2\,$km/s. The dotted curves show the hydrostatic
           solution, and the $\Gamma=1$ level. $\Gamma$ is arbitrarily
           enhanced (see text).}
  \vspace*{-1mm}
  \label{fig:dynamical}
\end{figure}

The first results of the dynamical models showed almost no mass loss
($\dot{M}\!\la\!10^{-10}\rm\,M_\odot/yr$), just some erratic
large-scale and long-term excursions for which the mass loss rate is
actually difficult to measure. We then approached rather extreme stellar
parameters ($M_{\star}\!=\!1\,M_{\odot}$, $T_{\star}\!=\!2500\,$K,
$L_{\star}\!=\!10000\,L_{\odot}$), still without success. Finally, in
order to see how a dust-driven wind could look like, we arbitrarily
enhanced the radiative acceleration by
\begin{equation}
  \Gamma = \Gamma_{\rm gas} + 5\,\Gamma_{\rm dust} \ .
\end{equation}
The results of this simulation is shown in Fig.~\ref{fig:dynamical}.

The models (also those without enhanced $\Gamma$) show extended warm
molecular layers, truncated by the $\Tg$-step described in
Sect.~\ref{sec:static} due to the line blanketing effect. The
pulsation of the star leads to a time-dependent extension of these
layers to roughly $1.5-2\,R_\star$. If dust forms, it fills in the gas
opacity gaps in frequency space which reduces the line blanketing
effect.

The formation of seed particles (see the nucleation rate $J_\star/\nH$ in
Fig.~\ref{fig:dynamical}) happens right above these molecular layers,
\ie very close to the star. The gas is cold here
($\Tg\!\approx\!700-900\,{\rm K}$) which is not revealed by models
using grey radiative transfer (Jeong\etal 2003, Ferrarotti\plus Gail 2006).

According to the model, two dust layers develop: almost pure glassy
{Al$_2$O$_3$} grains close to the star ($r\!\ga\!1.5\,R_\star$, even
partly inside the warm molecular layers up to
$\Tg\!\approx\!\Td\!\approx\!1500\,K$), and further out the more opaque
{Mg-Fe-silicates} grains which grow on top of the Al$_2$O$_3$ particles
with a very small, steadily increasing Fe content.

The temperature of the dirty dust grains is controlled by the
iron content, which has already been noted by Tielens et al.\ 
(1998)\nocite{twmj98}. The volume fraction of solid iron inclusions relaxes
quickly to a level where a further increase would cause too much
radiative heating and thus thermal re-evaporation of the solid iron
inclusions
\begin{equation}
  \Td \approx T_S^{\rm Fe}(\rho) \ ,
\end{equation}
where $T_S^{\rm Fe}$ is the sublimation temperature of solid
iron. This is a robust and recurrent result concerning many
simulations, because there is a stable self-regulation mechanism: The
iron content adapts quickly to any changes in the ambient medium
(density, radiation field) unless the density gets too small out in
the wind (here at $r\!\ga\!5\,R_\star$) where the degree of
condensation of Fe freezes in on a relatively low final level of
$\sim\!17\%$. It is this factor that kills the mass loss (in
comparison Mg: $\sim\!65\%$, Al: $\sim\!100\%$). Noteworthy,
silicates in AGB star winds are in fact observed to be Fe-poor
(Bowey\etal 2002)\nocite{bow2002}.

The radiative acceleration just even approaches the gravitational
deceleration around
$r\!\approx\!3.5-4\,R_\star$, which can be considered as an analog of
the sonic point in stationary winds. Determined by long-term averages
at the outer boundary, the mean mass loss rate, the mean outflow
velocity and the mean dust-to-gas ratio are
$\langle\dot{M}\rangle\!\approx\!2.3\times 10^{-9}\rm\,M_\odot/yr$,
$\langle v_\infty\rangle\!\approx\!2.6\,$km/s and $\langle \rho_{\rm
d}/\rho_{\rm g}\rangle\!\approx\!1.6\times 10^{-3}$, respectively (for
$\Gamma\!=\!\Gamma_{\rm gas}\!+\!5\,\Gamma_{\rm dust}$).

\section{Conclusions}

\begin{enumerate}

\item This letter reports on a {\it negative result}. Detailed
      dynamical models with frequency-dependent Monte Carlo radiative
      transfer and time-dependent formation of dirty dust grains
      cannot explain the observed magnitude of mass loss rates from
      oxygen-rich AGB stars, even in case of extreme stellar
      parameters (\ie high $L_\star/M_\star$ ratios).

\item The {\it role of solid iron and Fe-rich silicates} is crucial
      for the wind driving mechanism. These condensates are the only
      ones that are opaque around $1\,\mu$m and, thus, only these
      condensates can efficiently absorb the stellar light. Since the
      Fe containing condensates are not particularly stable, they form
      at too large distances from the star in order to provide an efficient
      mass loss mechanism.

\item Previous {\it grey models} (Jeong\etal 2003, Ferrarotti\plus
      Gail 2006) have calculated the radiation pressure on dust with
      Rosseland mean opacities which leads to a severe overestimation
      in the O-rich case, because the mid-IR dust absorption
      resonances are strongly peaked in a spectral region where the
      stellar flux is low.  The local flux $F_{\!\lambda}$ is not
      $B_\lambda(\Td)$-like, because the dust features are optically thick
      which makes the radiation field more isotrop and reduce
      $F_\lambda$ just where $\kappa_{\lambda,\rm\,ext}^{\rm\;dust}$
      is large.

 
\item The {\it dust condensation sequence} is strongly affected by
      radiative transfer effects. Pure, glassy condensates like
      Al$_2$O$_3$ have lower dust temperatures than solid Fe or
      Fe-rich silicates. The differences are as large as
      1000\,K, which favours the formation of the glassy condensates
      and prevents the formation of Fe-inclusions close to the
      star. The results in this paper are consistent with the
      observational finding of Al$_2$O$_3$ at radial distances as
      small as $1.5\,R_\star$ around $\alpha$\,Ori (Verhoelst\etal
      2006) as well as with the observed dust condensation sequence in
      O-rich AGB stars (Blommaert\etal 2006, Lebzelter\etal 2006),
      because Al$_2$O$_3$ can already exist in an extended atmosphere
      without mass loss, whereas the Mg-Fe-silicates form in the more
      distant wind regions which require mass loss.

\item The {\it mass loss mechanism} of oxygen-rich AGB stars and red
      supergiants is still a puzzle. Pulsations alone
      cannot drive an outflow because the radiative cooling of the gas
      is too efficient, even in non-LTE (Woitke 2003, Schirrmacher  et al.\ 
      2003)\nocite{woi2003,sws2003}. According to the results of this
      letter, even a combination of stellar pulsation and radiation pressure
      on dust is insufficient to drive the mass loss.  Do we have to
      re-visit Alfv{\'e}n-waves (\eg Vidotto\plus Jatenco-Pereira 2006)?
      \nocite{vp2006}

\end{enumerate}

\begin{acknowledgements}
  This work is part of the {\sc AstroHydro3D} initiative supported by
  the {\sc NWO Computational Physics programme}, grant
  614.031.017. The computations have been done on the parallel
  Xeon cluster {\sc Lisa} in Almere, the Netherlands, {\sc Sara} grant
  MP-103. The software used in this work was in part developed by the
  DOE-supported ASCI/Alliance Center for Astrophysical Thermonuclear
  Flashes at the University of Chicago.
\end{acknowledgements}

\vspace*{-4mm}

\end{document}